\documentstyle[preprint,aps]{revtex}
\begin{document}
\draft
\tighten
\preprint{LA-UR-93-4316}
\draft
\widetext
\title{
$\protect\protect\bbox{(j,0)\oplus(0,j)}$ Representation Space:
Majorana-Like Construct
\footnotemark[1]
\footnotetext[1]{This work was done under the auspices of the U. S.
Department of Energy. The lecture was delivered by D. V. Ahluwalia}
}
\author{D. V. Ahluwalia, T. Goldman, and  M. B. Johnson  }
\address{ Los Alamos National Laboratory,
Los Alamos, New Mexico 87545, USA}
\maketitle

\begin{abstract}
This is  second  of  the two  invited lectures presented at the ``XVII
International School of Theoretical Physics: Standard Model and Beyond' 93.''
The text is essentially based on a recent publication by the present authors
[Mod. Phys. Lett. A (in press)]. Here, after briefly reviewing the
$(j,0)\oplus(0,j)$ Dirac-like construct in the front form, we present a
detailed construction of the $(j,0)\oplus(0,j)$ Majorana-like fields.
\end{abstract}

\newpage

To facilitate the study of the massless limit, we work in the front-form
\cite{PAMD} Weinberg-Soper  formalism \cite{SW,DS}  recently developed in Ref.
\cite{MS}. The first lecture in this two-part series presented the {\it instant
form}  Dirac-like construction in the $(j,0)\oplus(0,j)$ representation space
and is not ideally suited to study the massless limit. The formalism that we
develop is valid for massive as well as massless particles.

\noindent{\bf 1.
Review of Dirac-Like $\protect\bbox{({\bf j},0)\oplus(0,{\bf j})}$ Spinors
in the Front Form}

The front-form   Dirac-like
$(j,0)\oplus(0,j)$ covariant spinors \cite{MS}  in the Weinberg-Soper formalism
(in   the chiral
representation) are defined as:
\begin{equation}
\psi\{p^\mu\}\,=\,\left[
\begin{array}{c}
\phi_R(p^\mu)\\
\phi_L(p^\mu)
\end{array}
\right]\quad .\label{spinor}
\end{equation}
The argument $p^\mu$ of chiral-representation spinors will be enclosed in curly
brackets $\{\,\,\}$. The Lorentz transformation  of the front-form $(j,0)$
spinors is given \cite{MS} by
\begin{equation}
\phi_R(p^\mu)\,=\,\Lambda_R(p^\mu)\,
\phi_R({\overcirc p}^\mu)\,=\,
\exp\left( \protect\bbox{ \beta\cdot {\bf J}}\right)\phi_R({\overcirc p}^\mu)\,
\quad,
\label{r}
\end{equation}
and  the front-form $(0,j)$ spinors transform as
\begin{equation}
\phi_L(p^\mu)=
\Lambda_L(p^\mu)\,
\phi_L({\overcirc p}^\mu)=\exp\left(-\protect\bbox{\beta^\ast
\cdot {\bf J}}\right)\,
\phi_L({\overcirc p}^\mu)
\quad. \label{l}
\end{equation}
The ${\overcirc p}^\mu$ represents the front-form four momentum for a
particle at rest: ${\overcirc p}^\mu \equiv (p^+=m,
\,p^1=0,\,p^2=0,p^-=m)\,$.  The ${\bf J}$ are the standard
$(2j+1)\times(2j+1)$ spin matrices, and ${\protect\bbox \beta}$ is the
boost parameter introduced in Ref. \cite{MS}
\begin{equation}
\protect\bbox{ \beta}
\,=\, \eta\,\left(\alpha\,v^r\,,\,\,-i\,\alpha\,  v^r\,,\,\,1\right)\quad,
\end{equation}
where
$
\alpha\,=\,\left[1\,-\,\exp(-\eta)\right]^{-1}\,,
$
 $v^r\,\,=\,\,v_x\,+\,i\,v_y$ (and
$v^\ell\,\,=\,\,v_x\,-\,i\,v_y$). In terms of the front-form variable
$p^+ \equiv E+p_z$, one can show that
\begin{equation}
\cosh(\eta/2)=\Omega
\left(p^+ + m\right)\,,\,
\sinh(\eta/2)=\Omega
\left(p^+ - m\right)\,,
\end{equation}
with
$
\Omega\,=\,\left[1/ (2 m)\right]\sqrt{m/ {p^+}}\,.
$
The norm $\overline{\psi}\{p^\mu\}\,\psi\{p^\mu\}$, with
\begin{equation}
\overline{\psi}\{p^\mu\}\,\equiv\,\psi^\dagger\{p^\mu\}\,\Gamma^0\,
,\quad\Gamma^0
\,\equiv\,
\left[
\begin{array}{cc}
0&\openone\\
\openone&0
\end{array}
\right]\quad,
\end{equation}
is so chosen that {\it  in the massless limit} :

\itemitem{a.} The Dirac-like $(j,0)\oplus(0,j)$
rest spinors identically vanish (there can be no massless particles at rest);
and
\itemitem{b.} Only the Dirac-like $(j,0)\oplus(0,j)$ spinors associated with
$h=\pm j$ front-form helicity \cite{MS} degrees of freedom survive. ($\openone$
is the $(2j+1)\times(2j+1)$ identity matrix.)

\medskip
\noindent
These requirements uniquely determine (up to a constant factor, which we choose
to be $1/{\sqrt{2}}\,$) the $(2j+1)$-element-column form of $\phi_R({\overcirc
p}^\mu)$ and $\phi_L({\overcirc p}^\mu)$ to be
\begin{equation}
\phi^R_{j}({\overcirc p}^\mu)\,=\,{m^j\over {\sqrt{2}}}
\left[
\begin{array}{c}
1\\
0\\
\vdots\\
0
\end{array}\right]
\,,\quad
\phi^R_{j-1}({\overcirc p}^\mu)\,=\,{m^j\over {\sqrt{2}}}
\left[
\begin{array}{c}
0\\
1\\
\vdots\\
0
\end{array}\right]
\,,\cdots\,\quad
\phi^R_{-j}({\overcirc p}^\mu)\,=\,{m^j\over {\sqrt{2}}}
\left[
\begin{array}{c}
0\\
0\\
\vdots\\
1
\end{array}\right]
\,\quad,\label{rests}
\end{equation}
[with similar expressions for $\phi_L({\overcirc p}^\mu)\,$] in a
representation in which $J_z$ is diagonal. The subscripts $h=
j,\,j-1,\,\cdots\,,-j$ on $\phi^R_h({\overcirc p}^\mu)$ in Eq. (\ref{rests})
refer to the front-form helicity \cite{MS} degree of freedom. The reader should
refer to Sec. 2.5 of Ref. \cite{MSM} for an alternate discussion of the
non-trivial nature (even though it appears as a ``normalization factor'')  of
the factor $m^j$  in Eq. (\ref{rests}).

\noindent{\bf 2.
Majorana-Like $\protect\bbox{({\bf j},0)\oplus(0,{\bf j})}$ Spinors
in the Front Form}

Following Ramond's work \cite{Ramond} on spin-$1\over 2$, we define the
front-form $(j,0)\oplus(0,j)$ {\it $\theta$-conjugate spinor }
\begin{equation}
\psi^\theta\{p^\mu\}\equiv \left[
\begin{array}{c}
\left(\xi\,\Theta_{[\,j]}\right)\,\phi^\ast_L(p^\mu) \\
\left(\xi\,\Theta_{[\,j]}\right)^\ast\,\phi^\ast_R(p^\mu)
\end{array}\right]\quad,\label{c}
\end{equation}
where $\xi$ is a c-number, and $\Theta_{[\,j]}$ is the Wigner's
time-reversal operator
(see Refs.: p. 61 of \cite{Ray}, Eqs. 6.7 and 6.8 of the first reference
 in \cite{SW},
 and Ch. 26 of \cite{EW}),
\begin{equation}
\Theta_{[\,j]}
\,\,{{\bf J}}\,\,\Theta_{[\,j]}^{-1} \,=\,-\,\protect\bbox{ {\bf
J}^{\,\ast}}\quad,
\end{equation}
and $^\ast$ denotes the operation of algebraic complex conjugation.
The parameter $\xi$ is fixed by imposing the constraint:
\begin{equation}
\left[\psi^\theta\{p^\mu\}\right]^\theta \,=\,\psi\{p^\mu\}\quad.
\label{cc}
\end{equation}
The time-reversal operator $\Theta_{[\,j]}$ is defined as:
$
\Theta_{[\,j]}\,=\,(-1)^{j+\sigma}\, \delta_{\sigma^\prime,\,-\sigma}\,.
$
It has the properties:
$
 \Theta_{[\,j]}^\ast\,\Theta_{[\,j]}\,=\,(-1)^{2j}\,,\,\,
\Theta_{[\,j]}^\ast\,=\,\Theta_{[\,j]}\,.
$
In the definition of $\Theta_{[\,j]}$, $\sigma$ and $\sigma'$ represent
eigenvalues of  ${\bf J}$. For $j={1\over 2}$ and $j=1$, the $\Theta_{[\,j]}$
have the explicit forms:
\begin{equation}
\Theta_{[1/2]}\,=\,\left[
\begin{array}{ccc}
0 & {}&-1 \\
1 & {} &0
\end{array}
\right]
\,,\,\,
\Theta_{[1]}\,=\,\left[
\begin{array}{ccccc}
0 &{}& 0 &{}& 1 \\
0 &{}& -1 &{}& 0 \\
1 &{}& 0 &{}& 0
\end{array}
\right]\quad.
\end{equation}
The  properties of %
Wigner's $\Theta_{[\,j]}$ operator  allow the parameter
$\xi$ involved in the definition of $\theta$-conjugation to be fixed as
$\pm\,i$ for fermions and $\pm \,1$ for bosons. However, without loss of
generality, we can ignore the {\it minus} sign [which contributes an {\it
overall} phase factor to the $\theta$-conjugated spinors
$\psi^\theta\{p^\mu\}$] and fix $\xi$ as
\begin{eqnarray}
\xi\,=\,\left\{\begin{array}{l}
i\,\,,\quad {\mbox{for fermions}} \\
1\,\,,\quad {\mbox{for bosons}}\qquad.
\end{array}\right. \label{alp}
\end{eqnarray}

The existence of  the
Majorana spinors for the $({1\over 2},0)\oplus(0,{1\over 2})$
representation space is
usually (see, e.g., p.16 of Ref. \cite{Ramond}) associated with the ``magic of
Pauli matrices,'' $\protect\bbox \sigma$. The reader may have already noticed
that
$i\,\Theta_{[1/2]}$  is identically equal to $\sigma_y$; and it is precisely
this  matrix that enters into the $CP$-conjugation of the
$({1\over 2},0)\oplus(0,{1\over 2})$ spinors.

The reason that the Majorana-like $(j,0)\oplus(0,j)$   representation
spaces, as opposed to the
$(j,0)\oplus(0,j)$ spaces spanned by
{\it Dirac-like spinors} defined by Eq. (\ref{spinor}),
can be constructed for arbitrary spins hinges upon two observations:
\medskip

\itemitem{1.} Independent of spin, the front-form boosts for the
$(j,0)$ and $(0,j)$ spinors have the property that \\
$\left[\Lambda_R(p^\mu)
\right]^{-1}\,=\,\left[
 \Lambda_L(p^\mu)\right]^\dagger\,,\quad
\left[\Lambda_L(p^\mu)\right]^{-1}\,=\,
\left[\Lambda_R(p^\mu)\right]^ \dagger
\,;
$ \\
and
\itemitem{2.} Existence of the Wigner's time-reversal matrix
$\Theta_{[\,j]}$
 for any spin.

\medskip
\noindent
These two observations, when coupled with the transformation
properties of the right- and left-handed  spinors, Eqs. (\ref{r},\ref{l}),
imply
that if $\phi_R(p^\mu)$ transforms as $(j,0)$, then
$(\zeta\,\Theta_{[\,j]})^\ast \,\phi_R^\ast(p^\mu)$ transforms as $(0,j)$
spinor. Similarly, if $\phi_L(p^\mu)$ transforms as $(0,j)$, then
$(\zeta\,\Theta_{[\,j]})^\ast\, \phi_L^\ast(p^\mu)$ transforms as $(j,0)$
spinor. Here, $\zeta=\exp(i\,\vartheta)$ is an arbitrary phase factor.  As
such,
we introduce $(j,0)\oplus(0,j)$ {\it Majorana-like spinors }
\begin{eqnarray}
(j,0) \,&\mapsto&\quad\rho\{p^\mu\}\,=\,
\left[
\begin{array}{c}
\phi_R(p^\mu) \\
\left(\zeta_\rho \,\Theta_{[\,j]}\right)^\ast\,\,\phi_R^\ast(p^\mu)
\end{array}\right]\quad,\nonumber\\
(0,j) \,&\mapsto&\quad \lambda\{p^\mu\}\,=\,
\left[
\begin{array}{c}
\left(\zeta_\lambda \,\Theta_{[\,j]}\right)\,\,\phi_L^\ast(p^\mu)\\
\phi_L(p^\mu)
\end{array}\right]
\,\,.\label{fb}
\end{eqnarray}
For formal reasons, the operator multiplying $\phi_L^\ast(p^\mu)$, in the
definition of $\lambda\{p^\mu\}$, is written as $\zeta_\lambda\,\Theta$ rather
than $\left(\zeta'_\lambda\,\Theta\right)^\ast$. %
What we have done, in fact, is to exploit 
the property $\Theta^\ast=\Theta$ and choose 
$\zeta_\lambda =
\zeta_\lambda^{'\,\ast}$. Since $\zeta_\lambda$ is yet to be determined, this
introduces no loss of generality. The advantage of all this is that
$\rho\{p^\mu\}$ and $\lambda\{p^\mu\}$ can now be seen as nothing but Weyl
spinors (in the $2(2j+1)$-element form)
\begin{equation}
\psi_R\{p^\mu\}\,=\,\left[
\begin{array}{c}
\phi_R(p^\mu)\\
0
\end{array}
\right]
\,,\,
\psi_L\{p^\mu\}\,=\,
\left[
\begin{array}{c}
0\\
\phi_L(p^\mu)
\end{array}
\right]
\quad,
\end{equation}
{\it added} to their  respective $\theta$-conjugates.
We now fix $\zeta_\rho$ and $\zeta_\lambda$ by demanding
(the defining property of the Majorana-like spinors):
\begin{equation}
\rho^\theta\{p^\mu\}\,=\,\pm\,\rho\{p^\mu\}\quad
{\mbox {and}}\quad
\lambda^\theta\{p^\mu\}\,=\,\pm\,\lambda\{p^\mu\}\quad,\label{sc}
\end{equation}
and find:
\begin{equation}
\zeta_\rho\,=\,\pm\,\xi \quad
{\mbox {and}}\quad
\zeta_\lambda\,=\,\pm\,\xi\quad.\label{zeta}
\end{equation}
The choice $\zeta_\rho =\zeta_\lambda=\,+\,\xi$ yields
self-$\theta$-conjugate spinors $\rho^{S_\theta}\{p^\mu\}$
and $\lambda^{S_\theta}\{p^\mu\}$;
while $\zeta_\rho =\zeta_\lambda=\,-\,\xi$ corresponds to
antiself-$\theta$-conjugate spinors
$\rho^{A_\theta}\{p^\mu\}$
and $\lambda^{A_\theta}\{p^\mu\}$.
The condition (\ref{cc})
is  satisfied not only by Majorana-like self-$\theta$-conjugate
spinors but also by antiself-$\theta$-conjugate spinors.

It may be noted that the Dirac-like spinors, Eq. (\ref{spinor}), and
the Majorana-like spinors, Eqs. (\ref{fb}), are two of the simplest
choices of spinors that
can be introduced in any $P$-covariant theory
\footnotemark[1]
\footnotetext[1]{A theory that is covariant under the operation of parity is
{\it not} necessarily a parity non-violating theory. See Sec. 7 for a brief
discussion of this point.}  in the $(j,0)\oplus(0,j)$ representation space. The
former describe particles with a definite charge (which may be zero), while the
latter are inherently for the description of  neutral particles.

Before we proceed further, we make a few observations on the definition of
$\theta$-conjugation. For the $({1\over 2},0)\oplus(0,{1\over 2})$ case, the
definition (\ref{c}) of $\theta$-conjugation can be verified to coincide with
$CP$-conjugation. The reader may wish to note that what Ramond (see Ref.
\cite{Ramond}, p. 20) calls a ``charge conjugate spinor,'' in the context of
spin-$1\over 2$,  is actually a $CP$-conjugate spinor. This, we suspect,
remains
true for fermions of higher spins also. Surprisingly, for the
$(1,0)\oplus(0,1)$ spinors (and presumably for bosons of higher spins also)
$\theta$-conjugation  equals $\Gamma^5\,C$ within a phase factor
\footnotemark[2]
\footnotetext[2]{ The charge conjugation
operator C, along with $P$ and $T$, for the $(1,0)\oplus(0,1)$ spinors and
fields was recently obtained in Ref. \cite{BWW}.}
of $-(-1)^{|h|}\,$.
The mathematical origin of
this fact may be traced to the constraint (\ref{cc}) and the property
$\Theta^\ast_{[j]}\,\Theta_{[j]}=(-1)^{2j}$ of %
Wigner's time-reversal
operator $\Theta\,$. One may ask if the  constraint (\ref{cc}) is changed
to
read $\left[\psi^\theta\{p^\mu\}\right]^\theta\,=\, -\psi^\theta\{p^\mu\}$ for
bosons, whether one can obtain an alternate definition of $\theta$-conjugation
(so that $\theta$-conjugation equals $CP$ for bosons also) to construct
self/antiself-$\theta$-conjugate objects. A simple exercise reveals that no
such construction yields self/antiself-$\theta$-conjugate objects. The reader
may wish to note parenthetically that when the
result (\ref{alp}) is coupled with
the definition of $\theta$-conjugation, Eq. (\ref{c}), we discover that the
operation of $\theta$-conjugation  treats the right-handed and  left-handed
spinors in a fundamentally asymmetric fashion for {\it fermions}. This is
readily inferred by studying the relative phases with which
$\left(\xi\,\Theta_{[\,j]}\right)\,\phi^\ast_L(p^\mu) $ and
$\left(\xi\,\Theta_{[\,j]}\right)^\ast\,\phi^\ast_R(p^\mu)$ enter in Eq.
(\ref{c}).

\noindent{\bf 3.
Explicit Construction of Majorana-Like $\protect\bbox{({\bf
1},0)\oplus(0, {\bf 1})}$ Spinors in the Front Form}

We now cast these formal considerations into more concrete form by studying the
$(1,0)\oplus(0,1)$ Majorana-like representation space   as an example. As in
Ref. \cite{MS}, we introduce
\footnotemark[3]
\footnotetext[3]{The generalized canonical representation
is introduced here for no other reason than 
to be able to compare the
results of the present work with our earlier work of Ref. \cite{MS}.}
the generalized canonical representation in the
front form:
\begin{equation}
\psi[\,p^\mu]\,=\,
{1\over{\sqrt{2}}}\,\left[
\begin{array}{ccc}
\openone &{\,\,\,}& \openone\\
\openone &{\,\,\,}& -\openone
\end{array}
\right]\,
\psi\{p^\mu\}\quad.\label{s}
\end{equation}
The argument $p^\mu$ of  canonical-representation spinors will be enclosed in
square brackets $[\,\,]$.
The boost $M(p^\mu)$, which connects the rest-spinors
$\psi[\,{\overcirc p}^\mu]$ with the spinors associated with front-form
four momentum $p^\mu$, $\psi[\,p^\mu]$, is determined from Eqs. (\ref{r}),
(\ref{l}), and (\ref{s}):
\begin{equation}
\psi[\,p^\mu]\,=\,
M(p^\mu)\,
\psi[\,{\overcirc p}^\mu]\,,\quad
M(p^\mu)\,=\,
\left[
\begin{array}{ccc}
{\cal A}&{\,\,}& {\cal B}\\
{\cal B}&{\,\,}& {\cal A}
\end{array}
\right]
\quad,
\end{equation}
with
$
{\cal A}= \Lambda_R(p^\mu)
+\Lambda_L(p^\mu)\,,\,\,
{\cal B}=
\Lambda_R(p^\mu)
-\Lambda_L(p^\mu)\,.
$

Using the identities (needed to evaluate $M(p^\mu)$ explicitly) given in Ref.
\cite{MS}, we first obtain the spin-$1$ $\rho^{S_\theta}[\,p^\mu]$ spinors. %
These are
tabulated in Table I. The $\rho^{S_\theta}[\,p^\mu]$ spinors satisfy the
following orthonormality relations: $ \overline\rho^{\,S_\theta}
_{h}[\,p^\mu]\,\rho^{S_\theta}
_{h'}[\,p^\mu]\,=\, m^2\Theta_{h h'} \,, $ where
\begin{equation}
\overline\rho_h[\,p^\mu]\,\equiv\,\left(\rho_h[\,p^\mu]\,\right)^\dagger
\Gamma^0\,,\quad \Gamma^0\,\equiv\,
\left[
\begin{array}{cc}
\openone & 0\\
0&-\openone
\end{array}
\right]\quad.
\end{equation}
The front-form $(1,0)\oplus(0,1)$ Majorana-like spinors, Table I,  should be
compared with the front-form $(1,0)\oplus(0,1)$ {\it Dirac-like} spinors
obtained in our recent work \cite{MS}. For instance,  in the massless limit for
the Dirac-like spinors,   the $h=\pm 1$ degrees of freedom are non-vanishing
and
the $h= 0$ degree of freedom identically vanishes. On the other hand, in the
massless limit, for the spin-$1$ Majorana-like spinors
$\rho^{S_\theta}[\,p^\mu]$, it is only the $h=+1$ degree of freedom that is
non-vanishing, while the $h=0$ and $h=-1$ degrees of freedom identically
vanish.

The origin of the above observation lies in the fact
\footnotemark[4]
\footnotetext[4]{Even though we make these observations for spin-$1$,
results similar to those
that follow are true for all spins (including spin-$1\over 2\,$).}
that the $(1,0)$ and $(0,1)$ boosts, $\Lambda_R(p^\mu)$ and $\Lambda_L(p^\mu)$,
 essentially become projectors of the $\phi^R_{+1}(p^\mu)$ and
$\phi^L_{-1}(p^\mu)$ as $m\to 0\,$.
To see this,  introduce

\begin{mathletters}
\begin{eqnarray}
{\cal Q}_R(m)&\,\equiv\,&
\left({m\over{p^+}}\right)\,\Lambda_R(p^\mu)\,=\
\left[
\begin{array}{ccc}
1 & 0 & 0\\
{\sqrt{2}}\,p^r/{p^+} & m/{p^+} & 0 \\
\left(p^r/p^+\right)^2 & \sqrt{2}\, m\, p^r /\left(p^+\right)^2 &
m^2/\left(p^+\right)^2
\end{array}
\right]\quad,\\
{\cal Q}_L(m)&\,\equiv\,&
\left({m\over{p^+}}\right)\,\Lambda_L(p^\mu)\,=\
\left[
\begin{array}{ccc}
m^2/\left(p^+\right)^2 & -\,\sqrt{2} \,m\, p_\ell /\left(p^+\right)^2
&\left(p^\ell/p^+\right)^2 \\
0 & m/p^+ & -\,\sqrt{2} \,p^\ell/p^+ \\
0&0&1
\end{array}
\right]\quad.
\end{eqnarray}
\end{mathletters}
The quasi-projector nature of ${\cal Q}_R(m\to 0)$ and
${\cal Q}_L(m\to 0)$ is immediately observed by verifying that:
${\cal Q}^2_R(m\to 0)={\cal Q}_R(m\to 0)$ and
${\cal Q}^2_L(m\to 0)={\cal Q}_L(m\to 0)\;$; but in general
${\cal Q}_R(m\to 0)+{\cal Q}_L(m\to 0) \not\to \openone\,$ and
${\cal Q}^\dagger_{R,L}(m\to 0)\not={\cal Q}_{R,L}(m\to 0)\,$.

To incorporate the $h=-1$ degree of freedom in the massless limit,
and to be able to treat the massive particles without introducing manifest
parity violation, we now repeat the above procedure for the
$\lambda^{S_\theta}[\,p^\mu]\,$
(and $\rho^{A_\theta}[\,p^\mu]$
and $\lambda^{A_\theta}[\,p^\mu]$  for the sake of completeness)
spinors. We find:
\begin{mathletters}
\begin{eqnarray}
&&\lambda_{-h}^{S_\theta}[\,p^\mu]\,=\,-(-1)^{|h|}\,
\rho_{h}^{S_\theta}[\,p^\mu]\quad,
\label{mmp}\\
&&\rho^{A_\theta}_h[\,p^\mu]\,=\,\Gamma^5\,\rho^{S_\theta}_h
[\,p^\mu]\,,\quad \Gamma^{5}\,=\,
\left[
\begin{array}{cc}
0 & \openone\\
\openone&0
\end{array}
\right]\quad,\label{extra}
\\
&&\lambda_{-h}^{A_\theta}[\,p^\mu]\,=\,(-1)^{|h|}\,
\Gamma^5\,\rho_{h}^{S_\theta}[\,p^\mu]\,=\,
(-1)^{|h|}\,\rho^{A_\theta}_h[\,p^\mu]
\quad;
\label{mmm}
\end{eqnarray}
\end{mathletters}
with $\openone=3\times 3$ identity matrix and
\begin{mathletters}
\begin{eqnarray}
&&\overline\rho^{\,S_\theta}_{h}[\,p^\mu]
\,\lambda^{S_\theta}_{h'}[\,p^\mu]\,=\, m^2\,\delta_{h h'}
\,=\,  \overline\rho^{\,A_\theta}_{h}[\,p^\mu]
\,\lambda^{A_\theta}_{h'}[\,p^\mu] \quad,\label{on}\\
&&\overline\rho^{\,A_\theta}_{h}[\,p^\mu]
\,\rho^{S_\theta}_{h'}[\,p^\mu]
\,=\, 0\,=\, \overline\rho^{\,S_\theta}_{h}[\,p^\mu]
\,\rho^{A_\theta}_{h'}[\,p^\mu] \quad.\label{onb}
\end{eqnarray} \end{mathletters}
As we will see in Sec. 6, the {\it bi-orthogonal} \cite{bio} nature of the
$\rho[\,p^\mu]$ and $\lambda[\,p^\mu]$ spinors results in a rather unusual
quantum field theoretic  structure for the $(1,0)\oplus(0,1)$ Majorana-like
field. Similar results hold true for other spins (including spin $1\over 2$).
The bi-orthogonal nature of the Majorana-like spinors is forced upon us by
self/antiself-$\theta$-conjugacy condition (\ref{sc})
 and cannot be changed as long as we require
that the basis spinors correspond to  definite spin projections (front form
helicity-basis in our case).

It should now be
recalled that for the Dirac-like $(1,0)\oplus(0,1)$ spinors
$u_\sigma\{p^\mu\}$ and $v_\sigma\{p^\mu\}\,$, we know \cite{BWW} from the
associated wave equation that the $u_\sigma\{ p^\mu\}$ spinors are associated
with the forward-in-time propagating solutions (the ``positive energy
solutions'')  $u_\sigma\{ p^\mu\}\,\exp[-i(E\,t-{ \bf p\cdot x})]$, and
$v_\sigma\{ p^\mu\}$ spinors are associated with the backward-in-time
propagating
\footnotemark[5]
 \footnotetext[5] {Recall that the usual
interpretation of the ``negative energy'' states as  {\it anti}particles fails
(see p. 66 of Ref. \cite{BH}) for bosons. On the other hand, the
St\"uckelberg-Feynman framework \cite{SF} applies equally to fermions and
bosons.}  solutions (the ``negative energy solutions'')
$v_\sigma\{p^\mu\}\,\exp[+i(E\,t- {\bf p\cdot x})]$. Can one infer similar
results by studying the wave equation associated with the front-form
$(1,0)\oplus(0,1)$ Majorana-like spinors$\,$?

\noindent{\bf 4. Wave Equation for Majorana-Like $\protect\bbox{({\bf 1},0)
\oplus(0,{\bf 1})}$ Spinors
in the Front Form}

Combining the Lorentz transformation properties for the $\phi_R(p^\mu)$ and
$\phi_L(p^\mu)$, given by Eqs. (\ref{r}) and (\ref{l}), with  the
definitions (\ref{fb}) of Majorana-like spinors,  we obtain the wave equations
satisfied by the $(1,0)\oplus(0,1)$ Majorana-like spinors. For the
$\rho\{p^\mu\}$ spinors, the wave equation we obtain reads (in chiral
representation, where it takes its simplest form):
\begin{equation}
\left[
\begin{array}{ccc}
-\zeta_\rho\,m^2\,\Theta_{[1]}&{\quad}&{\cal O}_1\\
{\cal O}_2&
{\quad}&-\zeta_\rho\,m^2\,\Theta_{[1]}
\end{array}
\right]
\rho\{p^\mu\}\,=\,0
\quad,
\label{majeq}
\end{equation}
where in the front form the operators   ${\cal O}_1$ and ${\cal O}_2$ are
defined as: $ {\cal O}_1\,=\, g_{\mu\nu}\,p^\mu\,p^\nu
\exp\left(-{\protect\bbox {\beta \cdot {\bf J}^\ast}}\right)\,\exp\left(
{\protect\bbox{ \beta^\ast\cdot  {\bf J}}}\right)\, $ and $ {\cal O}_2 =
g_{\mu\nu}\,p^\mu\,p^\nu \exp\left({\protect\bbox{ \beta^\ast\cdot {\bf
J}^\ast}}\right)\,\exp\left( - {\protect\bbox{\beta\cdot {\bf J}}}\right) \,.$
The non-zero elements of the front-form (the
flat space-time) metric $g_{\mu\nu}$
are: $g_{+-}={1\over 2}=g_{-+}$ and $ g_{11}=-1=g_{22}\,$. The wave equation
for the $\lambda\{p^\mu\}$ spinors is the same as Eq. (\ref{majeq})
 with $\zeta_\lambda$ being replaced by
$\zeta_\rho$.
The dispersion relations associated with the solutions of Eq.
(\ref{majeq}) are obtained by setting the determinant of the square bracket in
Eq. (\ref{majeq}) equal to zero. A simple, though somewhat lengthy, algebra
transforms the resulting equation into (true for all spin-$ 1$ Majorana-like
spinors, hence all reference to a specific spinor is dropped below): $
-\left(p^\ell\,p^r - p^+ p^- - \zeta\, m^2\right)^3\, \left(p^\ell\,p^r - p^+
p^- + \zeta\, m^2\right)^3 \,=\,0\,. $  As a result, the associated dispersion
relations read:
\begin{equation}
p^+\,=\,{ {p^\ell \,p^r\,+\zeta\, m^2}\over {p^-} }\,,\quad
p^+\,=\,{ {p^\ell \,p^r\,-\zeta \, m^2}\over {p^-} }\quad,
\end{equation}
each  with a multiplicity $3$ (for a {\it given} $\zeta$). Again, as seen in
Refs. \cite{DVAc,DVAf,SSB}, as in  the case for the Dirac-like
$(1,0)\oplus(0,1)$ spinors, the wave equation for the Majorana-like
$(1,0)\oplus(0,1)$ spinors contains tachyonic degeneracy. For the Dirac-like
$(1,0)\oplus(0,1)$ spinors, we find that  the tachyonic solutions can be
reinterpreted as physical solutions within the context of a quartic self
interaction and spontaneous symmetry breaking \cite{SSB}.  Here, we concentrate
on the physically acceptable dispersion relations $p^+\,=\,(p^\ell \,p^r\,+
m^2)/ {p^-}$; or equivalently $E^2\,=\,{\bf p}^{\,2} \,+\, m^2\,$.

The wave equation satisfied by the plane-wave solutions $ \rho\{x\}
=\rho\{p^\mu\}\,\exp(-i\epsilon\,p^\mu x_\mu)\, $ and $ \lambda\{x\}
=\lambda\{p^\mu\}\,\exp(-i\epsilon\,p^\mu x_\mu)\, $ is obtained by first
expanding the exponentials in Eq. (\ref{majeq}), in accordance with the
identities given in Appendix A of Ref. \cite{MS}, and then letting
$p^\mu\rightarrow i\partial^\mu$. Next, to determine $\epsilon$, %
we study the
resulting equation for the plane-wave solutions associated with  the rest
spinors. It is easily verified that for $\rho\{\,p^\mu\}$ as well as
$\lambda\{\,p^\mu\}$, it is not $\epsilon$ (directly)  but $\epsilon^2$ that is
constrained by the relation: $\epsilon^2= 1\,,$ giving $\epsilon=\pm 1\,$. This
is consistent with the intuitive understanding  in that we cannot distinguish
between the  forward-in-time propagating (``particles'')  and the
backward-in-time propagating (``antiparticles'') Majorana-like objects.
The above arguments are independent of which
representation we choose within the $(1,0)\oplus(0,1)$ representation space.

\noindent{\bf 5.
Majorana-Like $\protect\bbox{({\bf 1},0)
\oplus(0,{\bf 1})}$  Field Operator in the Front Form}

We now exploit the above considerations on the Majorana-like
spinors to construct the associated field operator. Generalizing the
spin-$1\over 2$ definition for a Majorana particle  of  Ref. \cite{BK},  we
define a general $(j,0)\oplus(0,j)$  Majorana-like field operator
$\Xi(x)$
\begin{equation}
U(C_\theta)\,\Xi(x)\,U^{-1}(C_\theta)
\,=\,\pm\,\Xi(x)
\quad. \label{u}
\end{equation}
In Eq. (\ref{u}), the ``$+$'' sign defines the self-$\theta$-conjugate and the
``$-$''  sign defines the antiself-$\theta$-conjugate field operator. The
explicit chiral-representation expression for $\theta$-conjugation operator
$C_\theta$ as contained in Eq. (\ref{c}) is
\begin{equation}
C_\theta\,=\,{\bf C}_\theta\,K\,=\,\left[
\begin{array}{cc}
0&\xi\,\Theta_{[\,j]}\\
\left(\xi\,\Theta_{[\,j]}\right)^\ast&0
\end{array}
\right]\,K\quad,
\end{equation}
where, $K$ complex conjugates (on the right) the objects in the Majorana-like
$(j,0)\oplus(0,j)$ representation space. For the example case of spin-$1$, when
Eqs. (\ref{u})  are coupled with the additional physical requirement that all
helicity degrees of freedom be treated symmetrically for manifest rotational-
and
$P$-covariance, the field operator $\Xi(x)$ is determined to be
\begin{mathletters}
\begin{eqnarray}
\Xi^{S_\theta}(x)  &=& \sum_{h=0,\pm1}\int {d^4 p}
{\Big[}{\cal
S}^{(\rho)}_{h}(p^\mu)\,\rho^{S_\theta}_{h}[\,p^\mu]\,\exp(-ip\cdot x)
\,+\,\eta_{_{GK}}\,
{\cal S}^{(\lambda)\,\dagger}
_h(p^\mu)\,\lambda^{S_\theta}_h[\,p^\mu]\,\exp(+ip\cdot x){\Big]}
\label{fos}\quad,\\
\Xi^{A_\theta}(x)  &=& \sum_{h=0,\pm1}\int {d^4 p}
{\Big[}{\cal A}^{(\rho)}
_{h}(p^\mu)\,\rho^{A_\theta}_{h}[\,p^\mu]\,\exp(-ip\cdot x)
\,+\,\eta_{_{GK}}\,
{\cal A}^{(\lambda)\,\dagger}
_h(p^\mu)\,\lambda^{A_\theta}_h[\,p^\mu]\,\exp(+ip\cdot
x){\Big]}\,,
\label{foa}
\end{eqnarray}
\end{mathletters}
where $\eta_{_{GK}}$ is the generalized
\footnotemark[6]
\footnotetext[6]{See footnote 19 of Ref. \cite{BK}.}
Goldhaber-Kayser phase factor;
and
\begin{mathletters}
\begin{eqnarray}
\left[{\cal S}^{(\rho)}_h(p^\mu)\,,\,\,{\cal S}^{(\rho)\,\dagger}
_{h'}(p^{\prime\,\mu})\right]&\,=\,&-\,(-1)^{|h|}\,
(2\pi)^3\,2 E(\vec p\,)\, \delta_{h,-h'}
\delta^3(\vec p- {\vec p}^{\,\prime})\quad,\label{coms}\\
\left[{\cal S}^{(\lambda)}_h(p^\mu)\,,\,\,{\cal S}^{(\lambda)\,\dagger}
_{h'}(p^{\prime\,\mu})\right]&\,=\,&-\,(-1)^{|h|}\,
(2\pi)^3\,2 E(\vec p\,)\, \delta_{h,-h'}
\delta^3(\vec p- {\vec p}^{\,\prime})\quad, \label{coma}
\end{eqnarray}\end{mathletters}
with similar expressions for the creation and annihilation operators of the
$\Xi^{A_\theta}(x)$ field. Several unusual features of expressions for the
field operators $\Xi(x)$, Eqs. (\ref{fos}) and (\ref{foa}), and commutators,
(\ref{coms}) and  (\ref{coma}), should be explicitly noted:
\medskip

\itemitem{ I.} The factor $-(-1)^{|h|} \,\delta_{h,-h'}$, rather than
the usual $\delta_{hh'}$, in the {\it rhs}
 of Eqs. (\ref{coms}) and (\ref{coma})
arises from the bi-orthogonal \cite{bio} nature of $\rho[\,p^\mu]$ and
$\lambda[\,p^\mu]$ spinors.
\itemitem{II.} The creation operator ${\cal S}^{(\lambda)\,\dagger}_h(p^\mu)$
for the plane wave $\lambda^{S_\theta}_h[p^\mu]\,\exp(ip\cdot x)$
is identical (within a
phase factor) to the
creation operator for the plane wave
 $\rho^{S_\theta}_{-h}[p^\mu]\,\exp(-ip\cdot x)$
\begin{equation}
{\cal S}^{(\lambda)\,\dagger}_h(p^\mu)\,=\,
-(-1)^{|h|}\,
{\cal S}^{(\rho)\,\dagger}_{-h}(p^\mu)\quad,
\end{equation}
\itemitem{} with similar comments applicable to
${\cal A}^{(\lambda)\dagger}_h(p^\mu)$
and
${\cal A}^{(\rho)\,\dagger}_h(p^\mu)$.

\medskip

\noindent
In addition, in view of our results of Sec. 4,  the association of the
$\rho[\,p^\mu]$ spinors with the forward-in-time propagating solutions and
$\lambda[\,p^\mu]$ spinors with backward-in-time propagating solutions in the
explicit expressions of $\Xi(x)$ above is purely a convention.

Finally, we wish
to emphasize that the field operators we arrive at differ from similar
expressions found in literature  for the $({1\over 2},0)\oplus(0,{1\over
2})$ Majorana field. Unlike the field operators $\Xi(x)$, these expressions
\footnotemark[7] \footnotetext[7]{See, for example, Eq. (3.25) of
Ref. \cite{BK}; and  Eq. (2.5) of Gluza and  Zralek's paper in Ref.
\cite{Zralek}.}
[even though they satisfy Eq. (\ref{u})] do not exploit the
Majorana-construction in the $(j,0)\oplus(0,j)$ representation space {\it and}
as a result cannot be expected to contain the full physical content of a truly
neutral particle.

\noindent{\bf 6.
Concluding Remarks}

We have succeeded in  extending the
Majorana-construction for the $({1\over 2},0)\oplus(0,{1\over 2})$
representation space to all $(j,0) \oplus(0,j)$  representation spaces  despite
the general impression that Majorana's original construction was due to a %
certain
``magic of Pauli matrices.'' We studied the $(1,0)\oplus(0,1)$ Majorana-like
representation space  in some detail and presented an associated wave equation.
Since nature has a host of neutral ``fundamental particles'' of spin-$ 1$ and
-$2$ and composite hadronic structures of even higher spins, the existence of
the Majorana-like $(j,0)\oplus(0,j)$ representation spaces introduced in this
work may have some physical relevance for the unification beyond the
electroweak theory and hadronic phenomenologies. In the massless limit,
$(j,0)\oplus(0,j)$ fields, independent of spin and independent of whether they
are Dirac-like or Majorana-like,  contain only two helicity degrees of freedom.
This observation allows the construction of higher-spin field theories without
introducing or imposing any auxiliary fields, negative-norm states, or
constraints. This fact may have some significance for theories involving
supersymmetric transformations, which transform between fermions and bosons,
and which are normally rife with non-physical, additional fields. It should be
explicitly noted that even though the construction of the $(j,0)\oplus(0,j)$
Majorana-like fields is {\it manifestly} covariant under parity, in general,
massive Majorana-like particles carry {\it imaginary} intrinsic parity
\cite{Parity} and hence these particles in interactions with
Dirac-like/Dirac particles (like charged leptons and quarks) naturally lead to
non-conservation of parity.

\noindent{\bf
Acknowledgements}
One of the authors (D.V.A.) warmly thanks Boris Kayser for a conversation, and
Otto Nachtmann and Marek Zralek for correspondence, on the  subject of Majorana
fields.

\nopagebreak %
{

\narrowtext
\begin{table}
\caption{Spin-$ 1$ self-$\theta$-conjugate
Majorana-like $\rho^{S_\theta}[\,p^\mu]$ spinors. Here $p^\pm=E\,\pm\,p_z$,
$p^r=p_x\,+\,i\,p_y$, and $p^\ell=p_x\,-\,i\,p_y$. The subscript $h=0,\,\pm 1$
on $\rho^{S_\theta}_h[\,p^\mu]$ refers to the {\it front-form helicity}
\protect\cite{MS} degree of freedom. The remaining spinors
$\lambda^{S_\theta}[\,p^\mu]$, $\rho^{A_\theta}[\,p^\mu]$, and
$\lambda^{A_\theta}[\,p^\mu]$ are related to  $\rho^{S_\theta}[\,p^\mu]$ via
Eqs. (\protect\ref{mmp}) to (\protect\ref{mmm}).} \begin{tabular}{ccc}
$\rho^{S_\theta}_{+1}[\,p^\mu]$ &$\rho^{S_\theta}_{0}[\,p^\mu]$
&$\rho^{S_\theta}_{-1}[\,p^\mu]$\\ \tableline ${1\over 2} \left[
\begin{array}{c} p^+\,+\,({p^\ell}^2/p^+)\\ \sqrt{2}\,(p^r\,-\,p^\ell) \\
p^+\,+\,({p^r}^2/p^+)\\ p^+\,-\,({p^\ell}^2/p^+)\\ \sqrt{2}\,(p^r\,+\,p^\ell)
\\ -\,p^+\,+\,({p^r}^2/p^+) \end{array} \right]$ & ${m \over\sqrt{ 2}} \left[
\begin{array}{c} p^\ell/p^+\\ 0\\ p^r/p^+\\ -\,p^\ell/p^+ \\ \sqrt{2} \\
p^r/p^+ \end{array} \right]$& $\,{m^2 \over 2} \left[ \begin{array}{c} 1/p^+\\
0\\ 1/p^+\\ -\,1/p^+ \\ 0\\ 1/p^+ \end{array} \right]$\\ \end{tabular}
\end{table}
}
\nopagebreak %

\end{document}